\documentclass[fleqn,usenatbib]{mnras}

\usepackage{newtxtext,newtxmath}

\usepackage[T1]{fontenc}

\DeclareRobustCommand{\VAN}[3]{#2}
\let\VANthebibliography\thebibliography
\def\thebibliography{\DeclareRobustCommand{\VAN}[3]{##3}\VANthebibliography}

\usepackage{graphicx}	
\usepackage{amsmath}	
\usepackage{amssymb}	

\title[Cosmological distribution of primordial black holes]
 {New evidence for a cosmological distribution of stellar mass
  primordial black holes}

\author[M. R. S. Hawkins]{
M. R. S. Hawkins $^{1}$\thanks{E-mail: mrsh@roe.ac.uk}
\\
$^{1}$Institute for Astronomy (IfA), University of Edinburgh,
 Royal Observatory, Blackford Hill, Edinburgh EH9 3HJ, UK\\}

\date{Accepted XXX. Received YYY; in original form ZZZ}

\pubyear{2021}

\begin{document}
\label{firstpage}
\pagerange{\pageref{firstpage}--\pageref{lastpage}}
\maketitle

\begin{abstract}
In this paper we show that to explain the observed distribution of
amplitudes in a large sample of quasar lightcurves, a significant
contribution from microlensing is required.  This implies the existence of
a cosmologically distributed population of stellar mass compact bodies
making up a large fraction of the dark matter.  Our analysis is based on
the lightcurves of a sample of over 1000 quasars, photometrically
monitored over a period of 26 years.  The intrinsic variations in quasar
luminosity are derived from luminous quasars where the quasar accretion
disc is too large to be microlensed by stellar mass bodies, and then
synthetic lightcurves for the whole sample are constructed with the same
statistical properties. We then run microlensing simulations for each
quasar with convergence in compact bodies appropriate to the quasar
redshift assuming a $\Lambda$CDM cosmology.  The synthetic lightcurve is
then superimposed on the amplification pattern to incorporate the effects
of microlensing.  The distribution of the resulting amplitudes can then be
compared with observation, giving a very close match.  This procedure does
not involve optimising parameters or fitting to the data, as all inputs
such as lens mass and quasar disc size come from independent observations
in the literature.  The overall conclusion of the paper is that to account
for the distribution of quasar lightcurve amplitudes it is necessary to
include the microlensing effects of a cosmologically distributed
population of stellar mass compact bodies, most plausibly identified as
stellar mass primordial black holes.

\end{abstract}




\section{Introduction}
\label{int}

The idea that dark matter in the form of compact bodies might be detected
by the microlensing of quasars was first suggested in a classic paper by
\cite{p73}.  They pointed out that if the Universe contains a roughly
critical density of compact bodies, then the probabilty that a distant
point light source such as a quasar will be gravitationally lensed is very
high.  For a lens mass of $\sim 1 M_\odot$ the distant source will be split
into two or more images with a separation of the order of $10^{-6}$ arcsec,
resulting in a change of observed integrated brightness.  In the event
that the lenses are of order a solar mass, the brightness will be observed
to change on a timescale of a few years as the lens passes across the
source as a result of their combined random motions.  \cite{c82} developed
this idea with particular reference to lenses of around a solar mass, and
concluded on the basis of datasets available at the time that such objects
were unlikely to be sufficiently numerous to close the Universe.

\begin{figure*}
\begin{picture}(300,200)(120,220)
\includegraphics[width=1.1\textwidth]{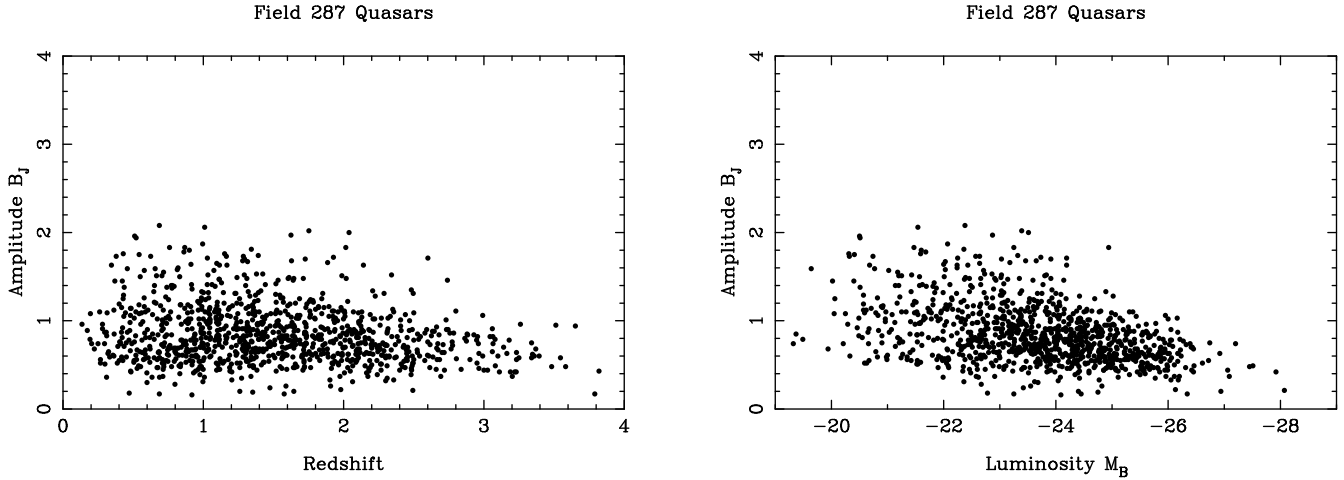}
\end{picture}
\caption{Amplitude as a function of redshift (left hand panel) and
 luminosity (right hand panel) for 1033 quasars from the Field 287
 survey.}
\label{fig1}
\end{figure*}

The idea that dark matter in the form of compact bodies might be detected
from the microlensing of quasars was investigated in more detail by
\cite{h93}, with the conclusion that a number of features in the
statistics of observed quasar lightcurves were consistent with the
predictions of microlensing, but were hard to explain in the context of
models of intrinsic variation current at the time.  Further evidence
favouring the microlensing of quasars came from features observed in
quasar lightcurves which were best explained as caustic crossing
events resulting from a large optical depth of microlenses \citep{h98}.
In the ensuing years a number of investigations in varied contexts
suggested the presence of dark matter in the form of compact bodies,
and are summarised in the form of a case for primordial black holes as
dark matter by \cite{h11}.

An alternative approach proposed by \cite{s93} to set limits on any
population of compact bodies involved simulating amplitude fluctuations
due to microlensing for varying combinations of lens mass and cosmological
density.  The amplitude distribution for each model was then compared with
the observed amplitudes for a sample of 117 quasar lightcurves covering a
period of 10 years \citep{hv93}.  The results of this study were that in
almost all cases the expected microlensing amplitudes from the simulations
were greater than those observed in the quasar sample.  However,
\cite{s93} pointed out that there were a number of assumptions associated
with these results.  These include the multiplication assumption for
calculating the amplification produced by adjacent lenses, the
distribution of tangential velocities for the lens relative to the source
and observer, and the size of the quasar disc.  Of these, the
multiplication assumption turned out to provide sufficient accuracy for
Schneider's purposes, the tangential velocities affect the estimation of
the lens masses which was not central to Schneider's line of argument, but
the most significant has turned out to be the source radius, which
Schneider set to $10^{15}$ cm or 0.4 lt-day.  Recent measures of the size
of quasar accretion discs from microlensing \citep{j15} and reverberation
mapping \citep{m18} imply values an order of magnitude larger than this,
with a characteristic radius of around 4 lt-day.  This has a major impact
on the results of \cite{s93}.

The approach pioneered by \cite{s93} to put bounds on the cosmological
density of compact objects was adapted by \cite{z03} to incorporate the
$\Lambda$CDM cosmology using the same sample of quasar lightcurves
\citep{hv93}, and to investigate the effects of source size and
tangential velocity on microlensing probabilities.  The assumption of a
nonzero value for the cosmological constant $\Lambda$ is well-known to
increase the optical depth to microlensing $\tau$ for a given cosmological
density $\Omega_L$ of compact bodies, or lenses.  This is well
illustrated by \cite{f90} for both point masses and isothermal spheres,
and has the effect of tightening the limits on the cosmological density
of lenses $\Omega_L$, as the probability of microlensing increases.  In
fact, \cite{f90} suggest that the analysis of the distribution of lenses
as a function of redshift could be a way of putting limits on the value of
the cosmological constant, but \cite{z03} opt to use a concordance value
for $\Lambda$, thus eliminating a potential free parameter in setting
limits on $\Omega_L$.  \cite{z03} also investigate the effect of changing
the assumptions on the transverse velocity distribution of the lenses,
which has the effect of changing the length of the simulated lightcurves
relative to the Einstein radius for given lens masses, and find that
constraints on the upper limit to $\Omega_L$ are only weakly affected by
such changes.  The most important result presented by \cite{z03} concerns
the effect of source size on limits to $\Omega_L$.  They conclude that
increasing the source radius $R_s$ to 4 lt-day results in no meaningful
constraint on the value of $\Omega_L$.  This can be seen in the context of
current measures of the accretion disc size for quasars of between 4 and
8 lt-day \citep{j15,m18}.

Given the difficulty of distinguishing between the intrinsic variability
of quasars and variations caused by microlensing, the emphasis for
detecting dark matter in the form of compact bodies switched to the halo
of the Milky Way.  The idea here was that if there were a significant
population of stellar mass compact bodies making up the dark matter
component of the Milky Way halo, then by monitoring several million
stars in the Magellanic Clouds one should occasionally observe a
microlensing event \citep{p86}.  This would be caused by a stellar mass
body crossing the line of sight to one of the Magellanic Cloud stars and
acting as a gravitational lens, amplifying the starlight in a
characteristic way as a function of time.  By analyzing the statistics of
the observed events, limits can then be put on the population of compact
bodies in the Milky Way halo, and hence on their fractional contribution
to the halo dark matter.  In order to put this idea to the test, the MACHO
collaboration \citep{a00} undertook a large scale photometric monitoring
programme in which they observed several million Magellanic Cloud stars in
two passbands on every available night for 6 years.  The final results
rested on their assumption of a high mass Milky Way halo, and the
reliability of their estimate of efficiency in detecting microlensing
events.  The widely quoted conclusion of this experiment was to rule out
a dark matter halo composed of solar mass compact bodies.  However, the 15
microlensing events which they did observe were far more than expected on
the basis of the stellar population of the Milky Way halo \citep{a00}.
Such a large population of halo stars would have been readily detected in
existing surveys for high velocity, low metalicity stars.  A detailed
analysis of the contribution of stars from the Milky Way and LMC to the
optical depth to microlensing is given by \cite{a00}.  This includes the
spheroid and disc populations of the Milky Way, and the halo population of
the LMC.  They find the total optical depth attributable to stellar
microlensing to be of order $\tau_* \sim 1 \times 10^{-8}$ compared
with the observed value of $\tau = 1.2 \times 10^{-7}$. The identity
of these lenses has yet to be determined, but they may indicate a less
simplistic solution to the nature of halo dark matter.  Other groups
carried out similar experiments, including EROS \citep{t07} and OGLE
\citep{w11}, which tended to tighten the constraints published by
\cite{a00}, although all groups used the same halo model for their
analysis.  After some 15 years, these result have now been widely
challenged.  \cite{h15} pointed out that more recent observations imply a
light halo for the Milky Way, which reduces the expected microlensing rate
such that it is compatible with a dark matter halo composed of stellar
mass compact bodies.  The paper also highlights a number of
inconsistencies in the way the detection efficiency was calculated.  Other
concerns have been raised by \cite{g17} who points out the sensitivity of
the result to the assumed mass function of the lenses, and \cite{c18} who
show that for a clumpy mass distribution in the halo, microlensing
constraints will become weaker especially in the stellar mass range.

More recently, a new window has been opened up for the detection of
black holes in the Galaxy \citep{w20}.  The idea here is to look in the
Galactic bulge for `dark' lensing events, where the contribution of
light from the lens is negligible.  This approach has already produced
some intriguing results, with a number of new black hole and neutron star
candidates.  Of particular interest is the failure to find useful evidence
for a mass gap between neutron stars with masses of up to $2 M_\odot$ and
black holes with masses over $5 M_\odot$.  This mass gap has been detected
in the mass distribution of X-ray binaries \citep{o10}, and there seems to
be no compellig reason why it is not seen in the Galactic bulge
observations.  There are several possible explanations, including `natal
kicks' where the black hole receives a boost to its tangential velocity at
formation, giving rise to misleading mass estimates.  An intriguing
possibility suggested by \cite{w20} is that the gap is filled by
primordial black holes, which would not be subject to the constraints from
X-ray observations as they would not be found in X-ray binaries.

\begin{figure*}
\centering
\begin{picture} (0,280) (255,0)
\includegraphics[width=0.49\textwidth]{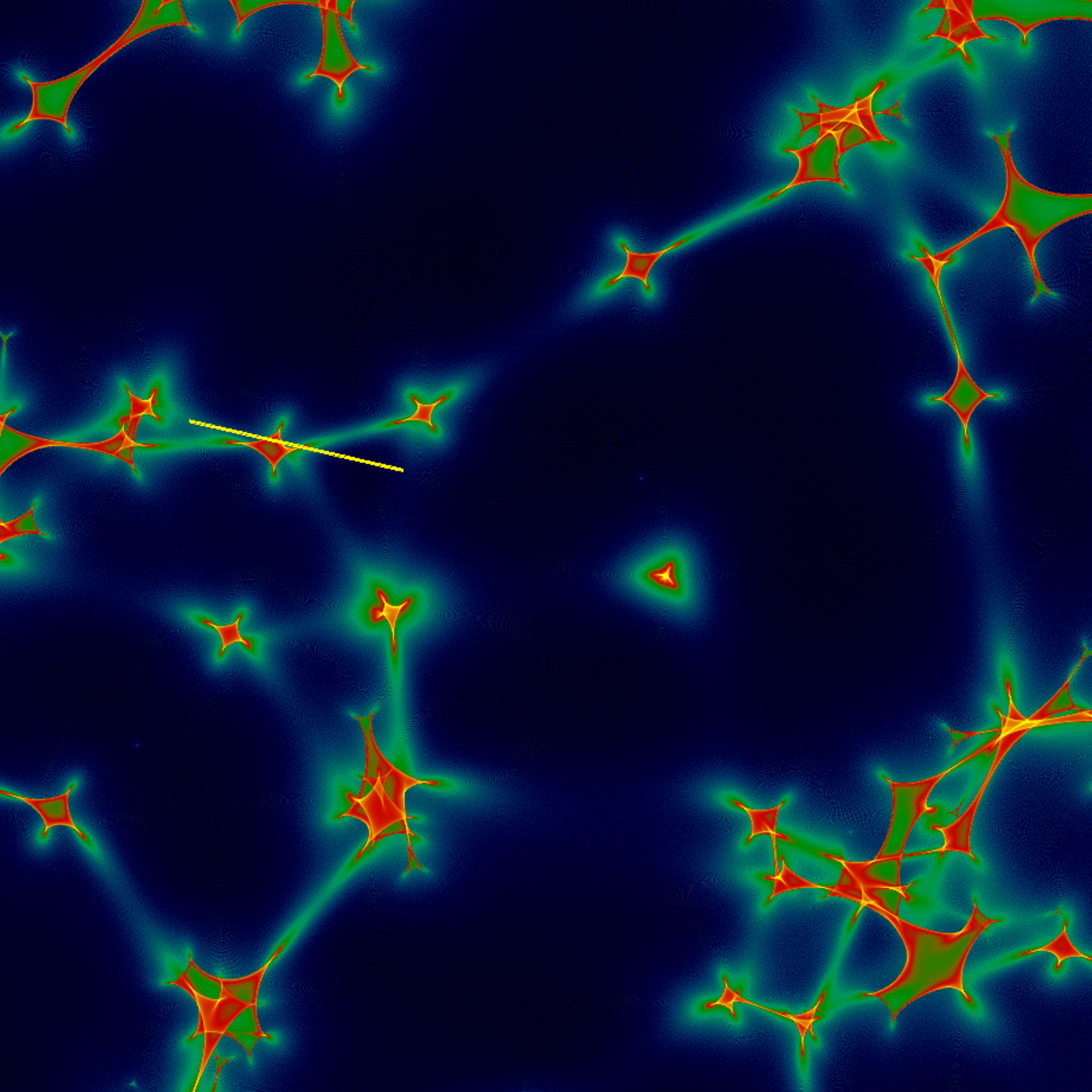}
\end{picture}
\begin{picture} (0,280) (-5,0)
\includegraphics[width=0.49\textwidth]{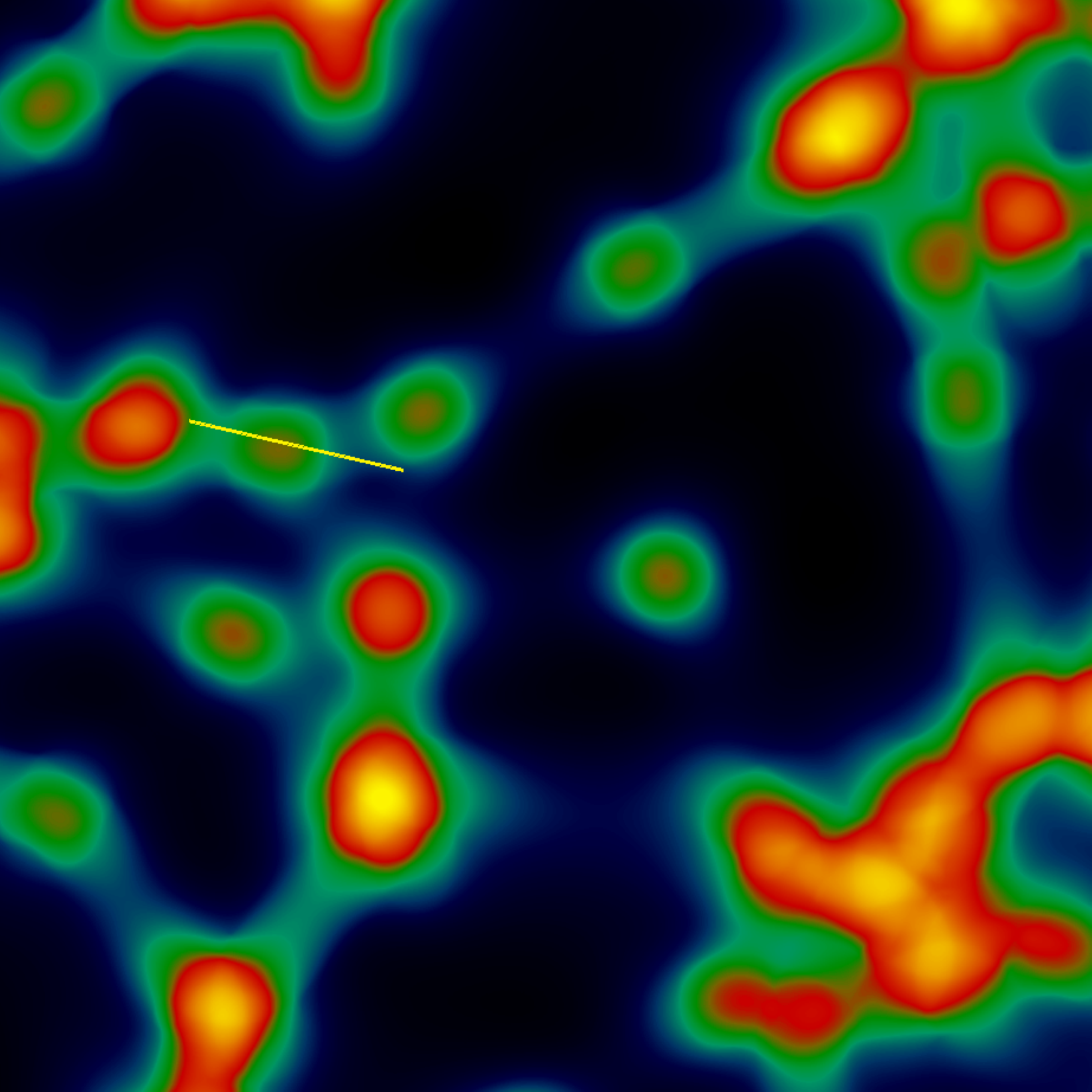}
\end{picture}
\caption{Microlensing magnification patterns for a population of 0.3
 M$_\odot$ bodies.  The frames have a side length of 16 Einstein radii in
 the source plane, and assume a source redshift $z = 2.5$ implying a total
 convergence $\kappa_* = 0.266$, and a most likely lens redshift
 $z = 0.82$ in a $\Lambda$CDM Universe with $\Omega_L = 0.3$.  In the left
 hand panel a point source is assumed, and the right hand panel is for a
 source of  half-light radius 4 lt-day.  Yellow lines indicate tracks
 across the amplification pattern corresponding to 26 years, the length of
 the lightcurves from the quasar monitoring programme.  A net transverse
 velocity of 600 km sec$^{-1}$ is assumed, and the position and
 orientation of the track is random.}
\label{fig2}
\end{figure*}

\begin{figure}
\begin{picture}(300,180)(10,30)
\includegraphics[width=0.55\textwidth]{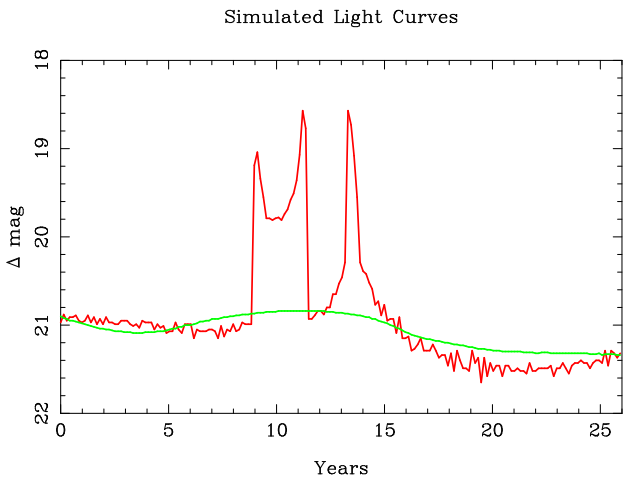}
\end{picture}
\caption{Lightcurves corresponding to the yellow tracks in
 Fig.~\ref{fig2}.  The red curve assumes a point source, and the green
 curve is for a source of half-light radius 4 lt-day, as for the left and
 right hand panels of Fig.~\ref{fig2} respectively.}
\label{fig3}
\end{figure}

The challenge to constraints on compact bodies in the Milky Way halo has
re-opened the possibility of detecting a population of compact bodies on
a cosmological scale that make up part or all of the dark matter.  A
promising place to look for such a population is in gravitational lens
systems where a quasar is split into multiple images by a massive galaxy
or cluster.  Although intrinsic variations in the quasar will show up in
the lightcurves of all the images, it is well known that the quasar
images also vary independently.  This is generally accepted to be the
result of microlensing along the line of sight to the quasar.  Until
recently, it has been believed that the microlenses are stars in the
halos of the lensing galaxy or cluster \citep{m09,p12}, but recent work
\citep{h20a,h20b} based on direct measures of starlight along the
line of sight to the quasar images has shown that the stellar populations
in the galaxy or cluster halos are too sparse to account for the observed
microlensing.  The compact bodies comprising the lenses must make up a
substantial proportion of the dark matter, and are tentatively identified
as primordial black holes.

In this paper we build on the result of \cite{s93} to determine whether a
cosmological distribution of compact bodies betray their presence in the
lightcurves of quasars which are not part of gravitational lens systems.
It is important to point out that as stars make up only around 1\% of
the critical density \citep{f98}, this implies an optical depth to
microlensing due to stars of $\tau_* \sim 0.01$.  This means that if
microlensing is detected at any significant level it must be associated
with a component of dark matter made up of compact objects.  We first
repeat Schneider's experiment in the context of the $\Lambda$CDM cosmology
with a sample of over 1000 quasars, and confirm the result of \cite{z03}
that if modern estimates of quasar accretion disc size are used, then the
range of amplitudes observed in quasar lightcurves is consistent with
microlensing simulations.  We then derive the distribution of intrinsic
amplitudes of variation by using as a template the lightcurves of luminous
quasars where the size of the accretion disc is too large to be
significantly microlensed by stellar mass bodies.  Combining this with the
distribution of amplitudes from microlensing simulations, we compare the
results with the observed distribution of amplitudes, and find that the
two distributions match each other closely.  None of the parameters used
in the modelling such as the quasar disc size or lens mass are optimized
or varied to fit the data, but come from independent and unrelated
measurements.

\section{Observations}
\label{obs}

The sample of lightcurves used by \cite{s93} and \cite{z03} covered 10
years and contained only 117 members \citep{hv93}, which severely limited
any investigation of statistical trends in redshift or luminosity.  Since
this early work the sample of lightcurves has been greatly enlarged, and
now contains data for 1033 quasars covering 26 years in the $B_J$ and 23
years in the $R$ band.  The data form part of a long term monitoring
programme of the ESO/SERC field 287 with the UK 1.2m Schmidt telescope at
Siding Springs Observatory in Australia from 1977 to 2002.  The plates
were measured by the SuperCOSMOS measuring machine at the Royal
Observatory Edinburgh \citep{h01} to give a range of parameters, including
instrumental magnitude, for each detected image. These magnitudes were
then calibrated with deep CCD photometric sequences to give true
magnitudes for each image.  For most years, 3 or 4 observations were
available, giving a mean magnitude for each year with an error of
$\sim 0.04$ mag.  This procedure is described in detail in \cite{h86,h03}
and references therein.  Examples of quasar lightcurves from the survey
are illustrated in \cite{h03}, which should give a feel for the quality of
the data.  In Fig.~\ref{fig1} we show plots of amplitude in the $B_J$
passband versus redshift and luminosity for the 1033 quasars in the
sample.  There is no obvious trend of amplitude with redshift, but a clear
decrease in amplitude with increasing luminosity.  This trend has been seen
in many studies (\cite{h00,v04} and references therein), but the
explanation for it has remained unclear.

\begin{figure*}
\begin{picture}(300,200)(120,230)
\includegraphics[width=1.1\textwidth]{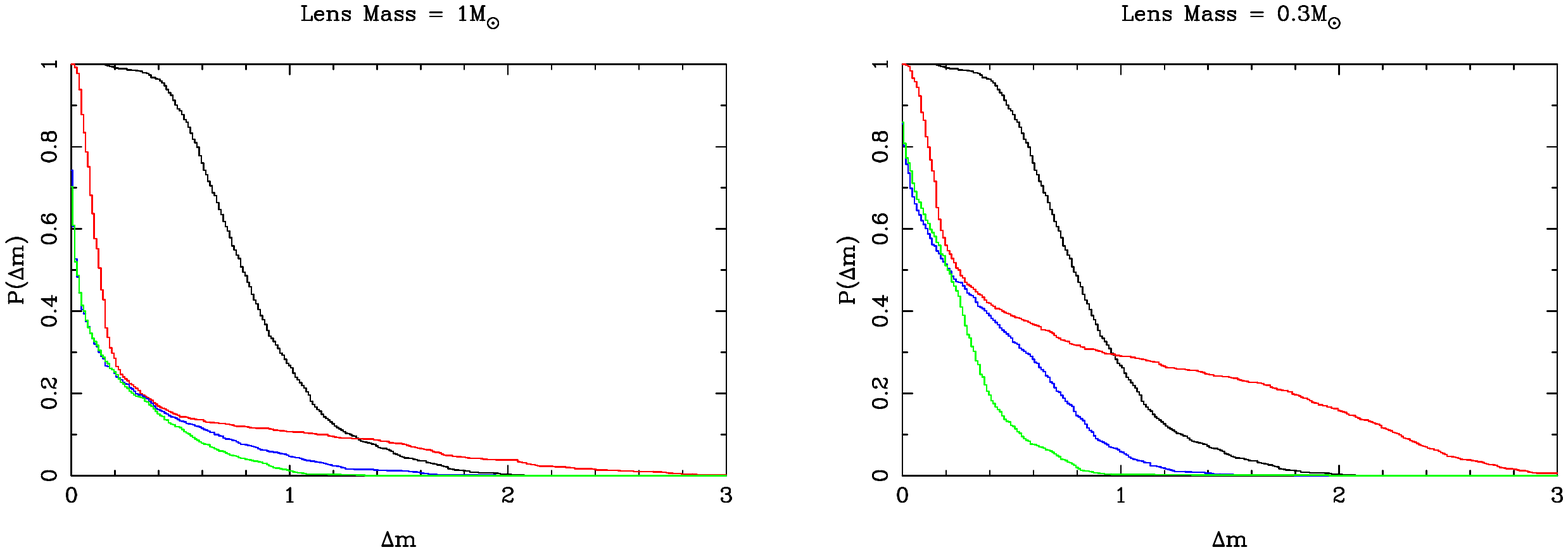}
\end{picture}
\caption{Cumulative probability $P(\Delta m)$ that a source varies by more
 than $\Delta m$ magnitudes for the sample of 1033 quasars from the Field
 287 survey for lens masses of 1.0 $M_\odot$ and 0.3 $M_\odot$.  The black
 curve represents the data, and the red, blue and green curves are
 microlensing probabilities for a point source and source radius of 4 and
 8 lt-day respectively.}
\label{fig4}
\end{figure*}

\section{Microlensing Simulations}
\label{micro}

The measure of variability used by \cite{s93} was the amplitude of the
lightcurve.  Other measures of variability such as rms variation are not
as well suited to analysing microlensing lightcurves, which are typically
characterized by sharp amplification events associated with caustic
crossings.  The purpose of Schneider's work was to see if these large
amplitude events were observed in quasar lightcurves, as a means of
putting limits on any cosmological distribution of compact bodies.  In
order to replicate the results of \cite{s93} with the new large sample of
quasar lightcurves we repeated his microlensing simulations, but taking
advantage of more recent knowledge of cosmological parameters.  The
simulations assume a $\Lambda$CDM cosmology with a set of cosmological
parameters reflecting tensions in current measurements.  For this purpose
we adopt the results of \cite{l16} who use a joint analysis of the
JLA+OHD+BAO datasets (see \cite{l16} for detailed references) to derive
a consistent set of cosmological parameters.  In particular, they find
$H_0$ = 67.8 kms$^{-1}$ Mpc$^{-1}$ and $\Omega_m = 0.350$ for the Hubble
constant and mass fraction respectively.  For the baryon density we use
the results of \cite{m20}, based on Big Bang nucleosynthesis, giving a
baryon fraction $\Omega_b = 0.048$.  Combining this with $\Omega_m$ we
obtain a value of $\Omega_L = 0.302$ for the cosmological density of
dark matter in the form of lenses.

Assuming this cosmology, the convergence $\kappa_*$ was calculated out to
the redshift of each quasar in turn, using the equations of \cite{f92}.
The software of \cite{w99} was then used to simulate an amplification
pattern with this value of $\kappa_*$.  The next step was to superimpose a
track in random position and orientation onto the simulation of the length
of the quasar lightcurve, using the conventionally assumed value for the
transverse velocity across the line of sight of 600 km sec$^{-1}$
\citep{k86}.  This required specifying the mass of the lenses, as the
Einstein radius $R_E$ provides a metric scale for the simulation.  On this
basis we adopted two representative masses of $1.0 M_\odot$ and
$0.3 M_\odot$.  The statistics of the amplitudes of the resulting
lightcurves were then compared with observations.  To illustrate the
results of the microlensing simulations, Fig.~\ref{fig2} shows a typical
amplification pattern for an assumed lens mass $M_L = 0.3 M_\odot$ and a
quasar of redshift $z = 2.5$.  This implies that the associated value of
$\kappa_*$, the convergence in compact bodies out to this redshift, is
$\kappa_* = 0.266$ \citep{f92}.  The left hand panel assumes a point
source, where the characteristic caustic pattern of microlensing
amplifications is clearly seen.  In the right hand panel a quasar disc of
half-light radius $R_s = 4$ lt-day is assumed, with the result that much
of the fine detail of the amplification pattern is blurred out, and the
overall amplification due to the lenses is much reduced.  The yellow line
shows a random track corresponding to the timespan of the quasar
lightcurves from the Field 287 sample.  The simulated microlensing
lightcurves are simply a regular sampling of amplifications along the
yellow tracks, with the distribution of lightcurve amplitudes to be
compared with observations.  In Fig.~\ref{fig3}, the lightcurve for the
point source in Fig.~\ref{fig2} is shown in red, and the cusp like
structure typically associated with caustic crossings is clearly visible.
The lightcurve for a source size $R_s = 4$ lt-day from the right hand
panel of Fig.~\ref{fig2} is shown in green, and illustrates the reduced
amplitude associated with a finite source size.

Fig.~\ref{fig4} shows the result of repeating the experiment illustrated
in Figure 6 of \cite{s93} with our new sample of quasar lightcurves.  To
summarize, for each quasar the value of $\kappa_*$ was calculated at the
redshift of the quasar for the $\Lambda$CDM cosmology of \cite{l16} with
$\Omega_L = 0.3$.  Microlensing simulations were then run using this value
of $\kappa_*$ for lens masses of $1.0 M_\odot$ and $0.3 M_\odot$, as
illustrated in Fig.~\ref{fig2}, and a track corresponding to the length of
the observed lightcurve superimposed as described above.  The simulated
microlensing amplitude for each quasar was then measured from its
lightcurve.  In Fig.~\ref{fig4} the normalised cumulative histogram of
amplitudes from the quasar sample is shown as a black line in both panels.
The coloured lines show the cumulative probability that a source varies by
more than $\Delta m$ magnitudes from cumulative histograms of amplitudes
from the microlensing simulations. The red line assumes a point source,
and the blue and green lines are for sources with radius 4 and 8 lt-day
respectively.  The choice of these accretion disc sizes is motivated by
the recent work of \cite{j15} and \cite{m18}.  It can be seen from both
panels in Fig.~\ref{fig4} that for a point source the predicted number of
large amplification microlensing events far exceeds those observed, but we
also find that for realistic source sizes of 4 and 8 lt-day the largest
predicted amplitudes are consistent with the observations.  This is in
agreement with the results of \cite{z03}.

The other notable feature of Fig.~\ref{fig4} is that microlensing only
appears to account for part of the total variability of the quasars.  This
is not unexpected as intrinsic variations are a well established feature
of AGN activity.  In the next Section we establish a procedure for
combining the varying luminosity of the quasars with the amplification due
to microlensing.

\section{Intrinsic quasar variability}

Distinguishing between intrinsic variations in quasar luminosity and the
effect of amplification by microlensing has proved to be a serious obstacle
to attempts to identify a population of compact bodies acting as
microlenses on a cosmological scale \citep{h96}.  For the quasar images in
gravitational lens systems this difficulty can be readily overcome and
microlensing amplifications identified \citep{h20a,h20b}, but in this
paper we are concerned with identifying microlensing amplifications in the
general cosmological population of quasars.  In order to characterize the
properties of the intrinsic variations of quasars we must look for a
regime where the variations are not significantly affected by
microlensing.  The right hand panel of Fig.~\ref{fig2} and the green curve
in Fig.~\ref{fig3} suggest that the variations due to microlensing of
quasars with large accretion discs will be small.  In Fig.~\ref{fig5} we
show the relationship between quasar disc size as determined from
reverberation mapping \citep{m18} and absolute magnitude $M_B$.  Also
shown is the best fit straight line with slope 0.5 tracing the expected
relation between log$(R_s)$ and $M_B$ for a constant surface brightness
disc.  This relation suggests that for quasars with $M_B < -25$ the size
of the accretion disc $R_s \gtrsim 6$ lt-day, rising to over 200 lt-day
for the most luminous quasars, and thus it may be concluded from
Fig.~\ref{fig3} that for such luminous objects microlensing will make a
negligible contribution to any variability.  On this basis we shall use
the amplitude distribution of luminous quasars ($M_B < -25$) as a template
for quasar intrinsic variation.  The black histogram in Fig.~\ref{fig6}
shows the amplitude distribution for quasars in the Field 287 sample with
$M_B < -25$, and is well fitted by a lognormal distribution

\begin{equation}
 f(x) = \frac{e^{-(\ln(x-\theta)-\mu)^2/(2\sigma^2)}}
 {(x-\theta)\sigma\sqrt{2\pi}}
\end{equation}

\noindent shown as a continuous blue line in Fig.~\ref{fig6}.  Also shown
in Fig.~\ref{fig6} as a red histogram is the lognormal function $f(x)$
rebinned for comparison with the data.  The best fit parameters are
$\sigma = 0.20$, $\theta = -0.36$ and $\mu = 0$, giving an adequate fit
with $\chi^2 = 3.39$ and 7 degrees of freedom.

In Fig.~\ref{fig7}, the black histogram shows the amplitude distribution
for the full Field 287 quasar sample, with no restriction on luminosity.
The continuous blue curve shows a lognormal distribution with the same
best fit parameters from Fig.~\ref{fig6}, but normalised to the 1033
quasars in the full Field 287 sample.  It will be seen that there is a
considerable excess of large amplitude variations in the histogram of
observed amplitudes which we tentatively identify as the result of
microlensing amplification by a population of compact bodies along the
line of sight to the quasars.  To test this hypothesis, we combine
intrinsic variations with microlensing amplification to give a
distribution of amplitudes for comparison with observations.  This
procedure is described in detail in the following Section~\ref{qlc}.

\section{Quasar lightcurve models}
\label{qlc}

\begin{figure}
\begin{picture}(300,180)(10,30)
\includegraphics[width=0.55\textwidth]{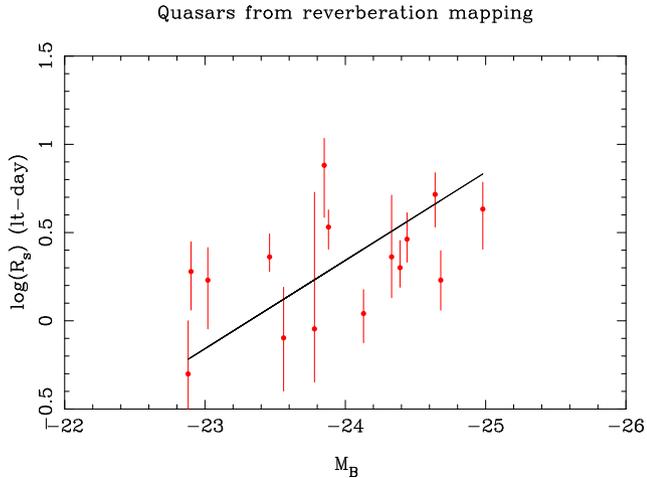}
\end{picture}
\caption{Relation between quasar half-light radius and absolute magnitude.
 Red filled circles are from reverberation mapping measures by
 Mudd et al. (2018).  The black line is a fit to the data assuming a
 constant surface brightness for the quasar disc.}
\label{fig5}
\end{figure}

\begin{figure}
\begin{picture}(300,180)(10,30)
\includegraphics[width=0.55\textwidth]{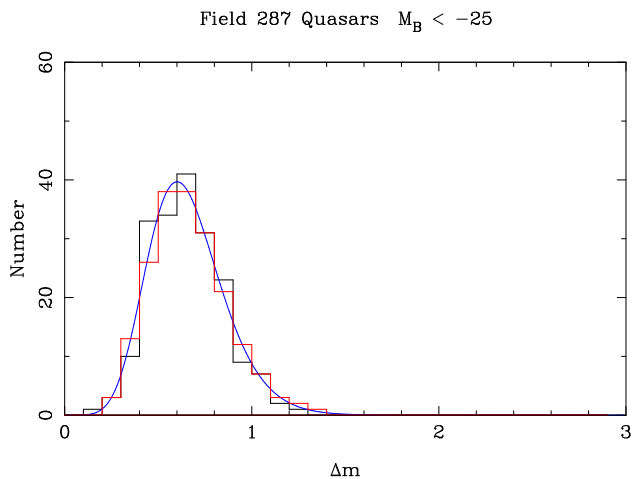}
\end{picture}
\caption{Histogram of amplitudes in the $B_J$ passband for the field 287
 sample of quasars with $M_{B} < -25$ (black line).  The blue line is the
 best fit lognormal distribution, and the red histogram shows the same
 data binned as for the observations.}
\label{fig6}
\end{figure}

\begin{figure*}
\centering
\begin{picture} (0,180) (255,30)
\includegraphics[width=0.55\textwidth]{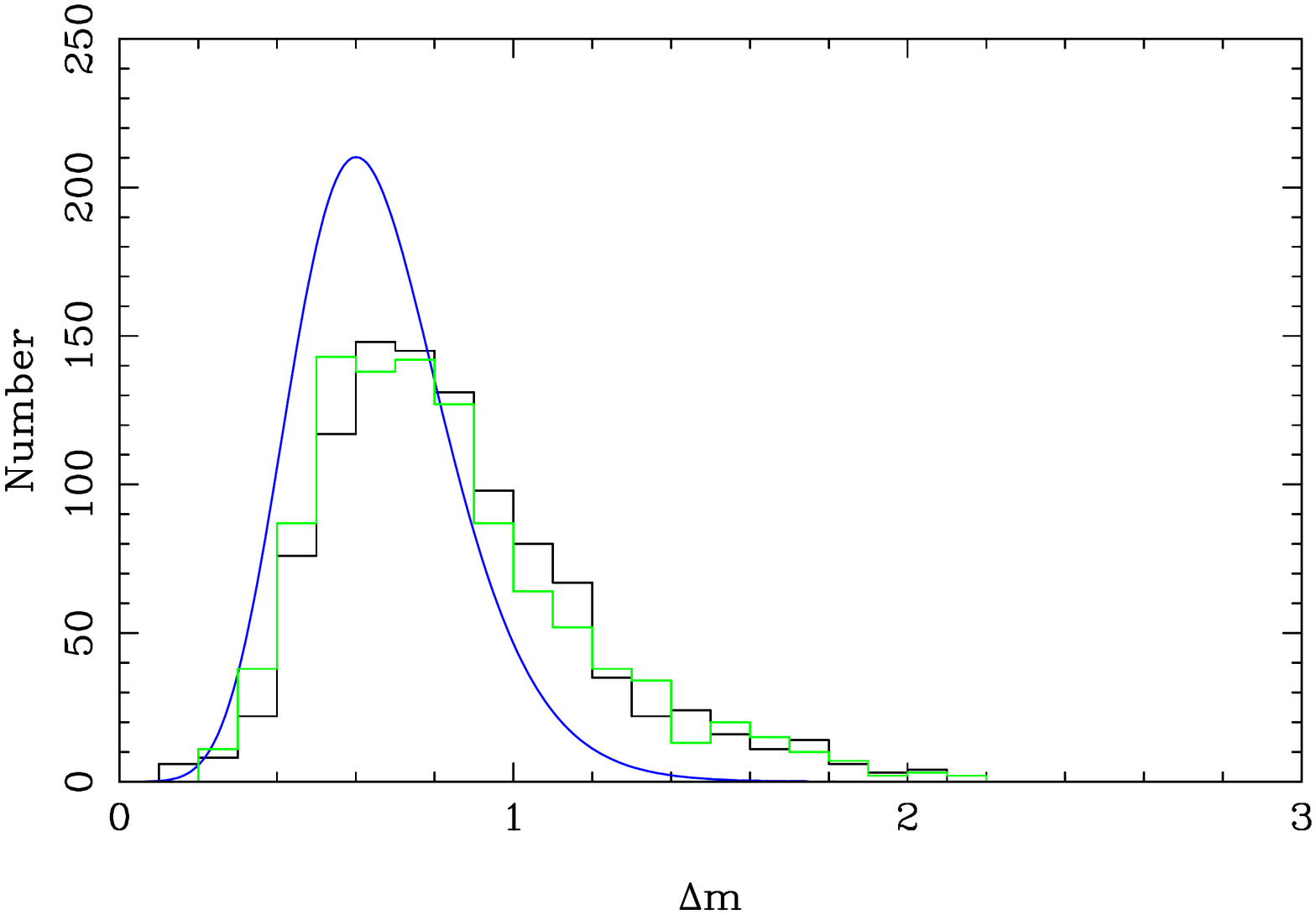}
\end{picture}
\begin{picture} (0,180) (-5,30)
\includegraphics[width=0.55\textwidth]{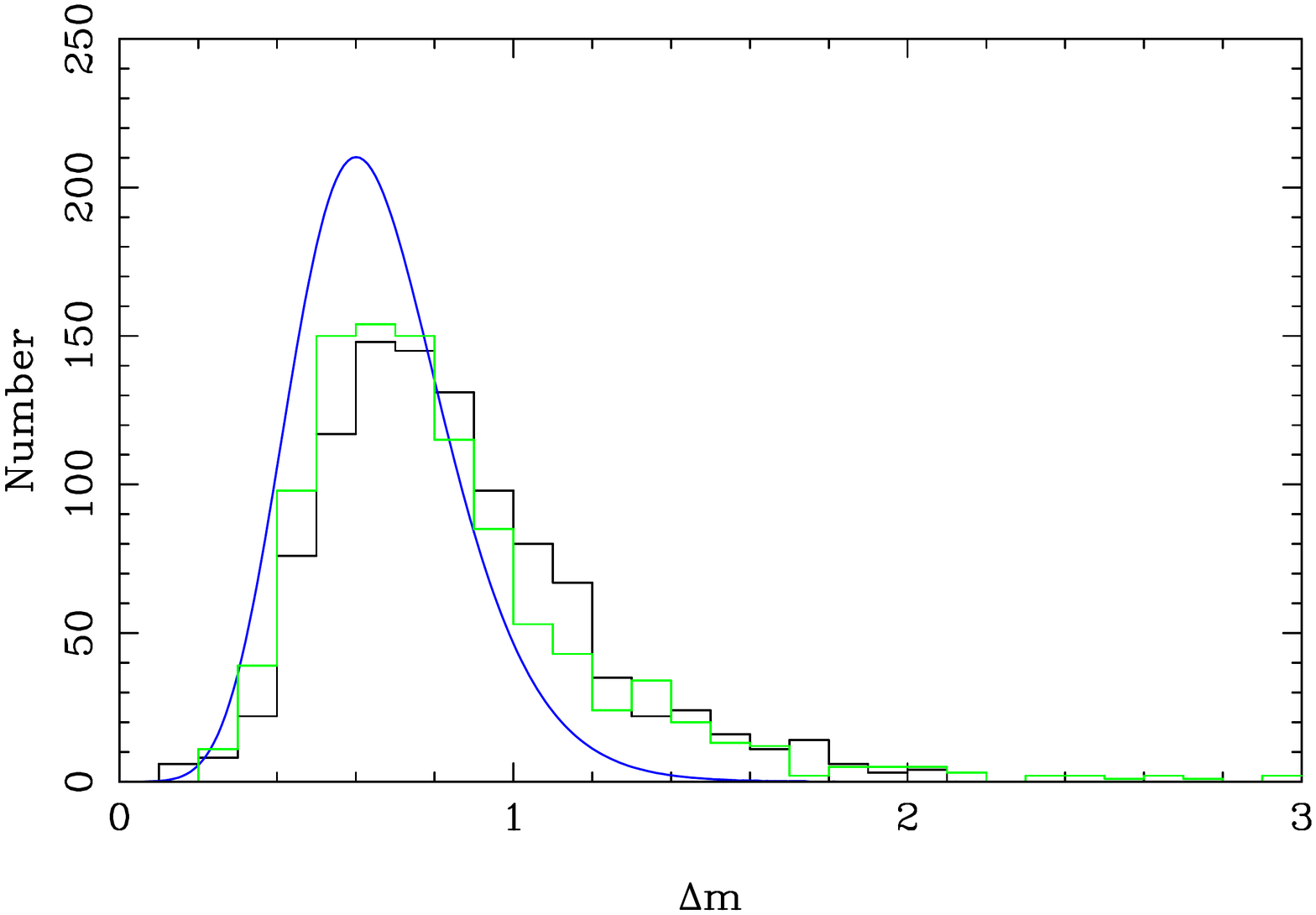}
\end{picture}
\caption{Histogram of amplitudes in the $B_J$ passband for the field 287
 sample of quasars (black line).  The blue line is the best fit lognormal
 distribution from Fig.~\ref{fig6}, normalised for the 1033 quasars in the
 full sample.  The green histogram shows the effect of amplification
 by microlensing based on the redshift of each of the sample members.  In
 the left hand panel a half-light radius of 4 lt-day is assumed for the
 quasar accretion disc, and in the right hand panel the size of the
 accretion disc is assumed to scale with luminosity, as illustrated in
 Fig.~\ref{fig5}.}
\label{fig7}
\end{figure*}

In order to model the intrinsic variations of the quasar sample we first
allocate to each quasar in turn an amplitude chosen at random from the
1033 amplitudes in the normalised lognormal distribution illustrated by
the blue line in Fig.~\ref{fig7}.  This ensures that the amplitudes of
the intrinsic variations accurately follow a lognormal distribution
function with parameters from the fit to the data in Fig.~\ref{fig6}.
We then construct synthetic lightcurves normalised to the allocated
amplitude from a power law spectrum \citep{h07}, assuming random phases.
The idea behind creating a lightcurve in this way is to be able to model
the effect of amplification by microlensing on the intrinsic brightness
changes of the quasar as it traverses the caustic pattern associated with
a cosmological distribution of compact bodies comprising the dark matter.

To model the effect of microlensing on these lightcurves of intrinsic
variations, the procedure described in Section~\ref{micro} was used to
simulate a lightcurve for each quasar at the corresponding redshift,
assuming a characteristic mass for the lenses of $0.3 M_\odot$.  The
amplification at each epoch of the microlensing lightcurve was then
applied to the corresponding point of the lightcurve of intrinsic
variations to model the total change in brightness of the quasar.  The
amplitude of this combined lightcurve was then measured to give a
statistical estimate of the observed quasar amplitude in the Field 287
sample.  The green histogram in the left hand panel of Fig.~\ref{fig7}
shows the simulated distribution of amplitudes assuming a quasar disc
half-light radius $R_s = 4$ lt-day.

The green histogram of simulated amplitudes in the left hand panel of
Fig.~\ref{fig7} closely follows the histogram of observed amplitudes in
black.  It is important to point out that the simulations are not fitted
to the data by optimizing any free parameters, but rest on three
assumptions, independently supported by observation.  The first is the
choice of $0.3 M_\odot$ for the characteristic lens mass.  This choice is
motivated by the results of the MACHO collaboration \citep{a00}, discussed
in Section~\ref{int}, who detected a large unidentified population of
compact bodies of around $0.3 M_\odot$ in the Galactic halo, and which are
seen as plausible dark matter candidates.  The second assumption is that
the photometric variations of luminous quasars are not significantly
affected by microlensing amplifications due to the large size of the
accretion disc, and thus provide a good template for intrinsic
variability.  The third assumption is the size of the quasar accretion
disc.  In the first instance we assume $R_s = 4$ lt-day, based on
microlensing \citep{j15} and reverberation mapping \citep{m18} measures,
and the green histogram in the left hand panel of Fig.~\ref{fig7} is based
on this assumption.  An alternative approach is to assume that the surface
brightness of a quasar disc is constant, and hence the disc radius $R_s$
scales with absolute magnitude as suggested by Fig.~\ref{fig5}.  The green
histogram in the right hand panel of Fig.~\ref{fig7} is based on this
assumption, and it is interesting to note that the overall shape is little
changed.

\begin{figure}
\begin{picture}(300,180)(10,30)
\includegraphics[width=0.55\textwidth]{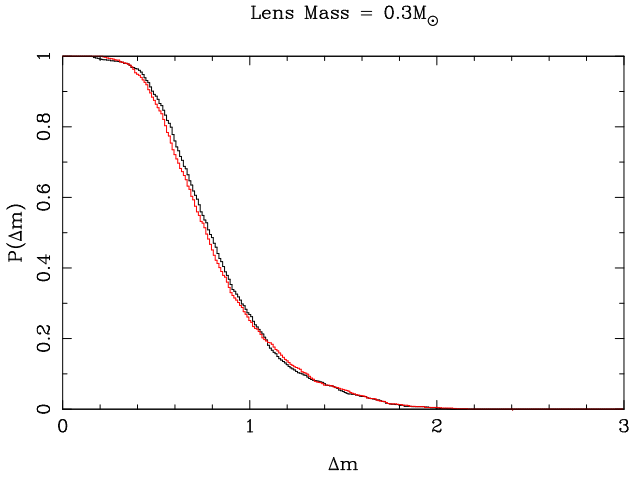}
\end{picture}
\caption{Cumulative probability $P(\Delta m)$ that a source varies by more
 than $\Delta m$ magnitudes for the sample of 1033 quasars from the Field
 287 survey for lens mass $0.3 M_\odot$.  The black curve represents the
 data, and the red curve shows the cumulative probability for a combination
 of lognormal intrinsic variation combined with microlensing amplifcation
 assuming a quasar disc radius $R_s = 4$ lt-day.}
\label{fig8}
\end{figure}

We are now in a position to repeat the cosmological test introduced by
\cite{s93} using a much larger sample of quasar lightcurves, spanning a
much longer time period.  Rather than assuming a constant quasar
luminosity, we have now incorporated intrinsic changes in quasar
luminosity based on photometric variations in the most luminous
quasars where the size of the accretion disc makes microlensing
amplification negligible.  In Fig.~\ref{fig8} we show an updated version
of Schneider's cumulative probablility plots where intrinsic variations
are amplified by microlensing from a cosmological population of stellar
mass lenses with a dark matter distribution. The combination of intrinsic
and microlensing variations provides a good match to the observations, and
we discuss the implications of this in the following Section.

\section{Discussion}

\begin{figure}
\begin{picture}(300,180)(10,30)
\includegraphics[width=0.55\textwidth]{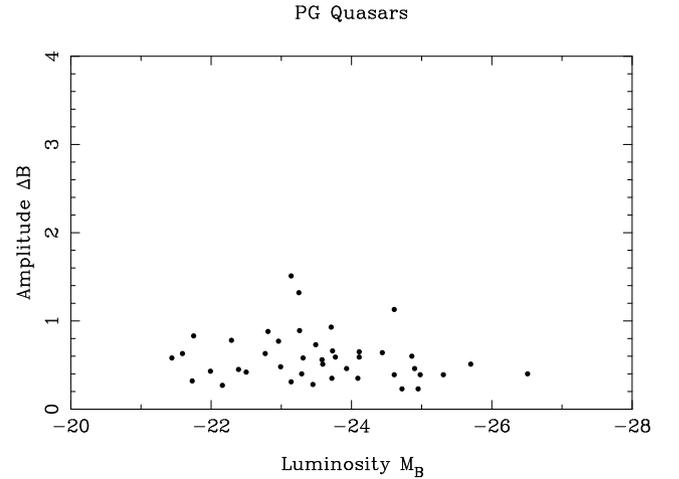}
\end{picture}
\caption{Amplitude as a function of luminosity for the 42 low redshift
 quasars in the Palomar-Green survey from Giveon et al. (1999).}
\label{fig9}
\end{figure}

The investigation behind this paper was prompted by the detection of a
large population of compact solar mass bodies from microlensing
observations in two separate environments.  In the first case \citep{a00}
the lenses were located in the dark halo of the Milky Way, and in the
second case the lenses were observed in the halos of distant massive
galaxies and a galaxy cluster \citep{h20a,h20b}.  These projects are
briefly described in Section~\ref{int}.  An important aspect of both
these results was that the population of stellar mass compact bodies
required to account for the observed microlensing was not consistent
with any plausible population of stars.  In the case of the Milky
Way, inventories of the stellar populations in the spheroid and disc have
been painstakingly built up over many decades from direct star counts, and
from kinematic measures from spectroscopy and proper motions
\citep{r86,g89}.  These data have been supplemented by photometric and
spectrometric abundance measurements to give an accurate overall map of
the distribution of stellar populations in the Milky Way, with no room for
a large new population of stellar microlenses.  Early work on stellar
populations of other galaxy types \citep{v75} is consistent with this
picture, and the extensive work done since then on the structure of
galaxies has found no evidence for such unexpected populations of stars.

The second environment where a large population of solar mass compact
bodies have been detected is in gravitational lens systems where a massive
galaxy is seen to split a quasar into two or four separate images.  It is
well established that intrinsic variations in the quasar are seen in the
individual images, separated by a small time difference reflecting the
time for the quasar light to reach each image.  More importantly for our
purposes, the individual quasar images are also seen to vary
independently.  This is widely recognised as the result of microlensing,
as the the light from the quasar follows different trajectories to each
image, and hence encounters different amplification patterns
from any distribution of compact bodies along the light path.  The
timescale of the microlensing events implies the lenses are around a
stellar mass, and one might conclude from these observations that this is
evidence for a cosmologically distributed population of stellar mass
compact bodies.  Given the ubiquity of the microlensing amplifications
these compact bodies would make up a large fraction of the dark matter.
There is however an important caveat with this conclusion.  By its very
nature, in a gravitational lens system a massive galaxy will lie along
the line of sight to the quasar, with the possibility that the quasar
images will lie close enough to the outskirts of the lensing galaxy for
the galactic stars to act as microlenses.  This possibility has been
examined in detail for a small sample of gravitational lens systems where
the microlensed quasar images appear to lie well clear of the stellar
population of the lensing galaxies \citep{h20a}.  Surface brightness
measures from from Hubble Space Telescope frames in the infrared were
converted to surface mass density, and hence optical depth to microlensing
$\tau$ to measure the probability both analytically and from computer
simulations of stellar microlensing amplification.  The results showed
that the probability that the observed microlensing could be caused by
stars was of the order of $10^{-4}$.  A similar analysis was carried out
for the cluster lens SDSS J2004+4112 where the micolensed images lie some
60 kpc from the cluster centre, and the probability of microlensing by
stars again appears to be negligibly small \citep{h20b}.

On the basis of these constraints on microlensing by stars, the population
of compact bodies making up the microlenses were identified as a component
of the dark matter, most plausibly in the form of stellar mass primordial
black holes \citep{h20a}.  Given the detection of such a population of
compact bodies in the halo of the Milky Way and more distant galaxies, it
has been the purpose of this paper to look for evidence of their presence
in the form of a cosmological distribution, tracing out the dark matter.

As a starting point for the investigation we have taken the cosmological
test proposed by \cite{s93} to put limits on the extent to which compact
bodies can make up a component of the dark matter.  Using a sample of
quasar lightcurves a factor of 10 larger than that available to
Schneider, and covering a timespan increased from 10 to 26 years,
we repeated his simulations in the context of the $\Lambda$CDM cosmology.
The results illustrated in Fig.~\ref{fig4} broadly speaking confirm
Schneider's claim that for a point source the predicted large amplitude 
microlensing events are not seen in the observations.  However, in line
with the results of \cite{z03} we find that for a quasar disc radius
$R_s \gtrsim 4$ lt-day there is no such conflict between microlensing
simulations and observations.

The next step in the programme was to model the intrinsic variations of
the quasars.  To do this we made use of the relation between quasar disc
radius $R_s$ and luminosity from \cite{m18} as illustrated in
Fig.~\ref{fig5}, to identify quasars with sufficiently large accretion
discs that microlensing amplifications would be negligibly small.
This was achieved by adopting a limit of $M_B < -25$, to create a sample
of luminous quasars with implied disc radii $R_s$ ranging from 6 to over
200 lt-day.  The lightcurves in Fig.~\ref{fig3} give an idea of the extent
to which microlensing amplitudes are reduced as $R_s$ increases.  On this
basis the observed distribution of lightcurve amplitudes is a measure of
intrinsic variation, and is well-fitted by a lognormal distribution.
Combining these intrinsic variations with microlensing amplifications
gives an excellent match to the observed amplitude distribution, as
illustrated in Fig.~\ref{fig7}.  Finally, replotting these data as a
cumulative distribution as for the cosmological test proposed by
\cite{s93} results in a distribution of amplitudes very close to the
observations.

Apart from the adoption of the $\Lambda$CDM cosmology which is not
controversial in cosmological studies, this final result rests on the
three assumptions outlined in Section~\ref{qlc}.  The unidentified
population of compact bodies  detected by \cite{a00} were found to have a
characteristic mass of around $0.3 M_\odot$, which is the the same as the
mass used to account for the observed microlensing in galaxy halos
\citep{m09,p12} and clusters \citep{h20b}.  It is also close to the
preferred mass of $\sim 0.7 M_\odot$ from  theoretical studies of
primordial black hole formation during the QCD phase transition
\citep{b18,c21}.

The reduction in microlensing amplitude with increasing source size was
thoroughly discussed and modelled some time ago by \cite{r91}, and is
again illustrated here in Fig.~\ref{fig3}.  It is clear from
Fig.~\ref{fig5} that quasars with luminosity $M_B \lesssim -25$ already
have an accretion disc radius $R_s \gtrsim 6$ lt-day, half as large again
as the assumed value for $R_s$ in Fig.~\ref{fig3} which already shows
much reduced variation due to microlensing.  On this basis we have assumed
that the variations in quasars with $M_B \lesssim -25$ are intrinsic
variations in quasar lumininosity.

The third assumption that the size of the quasar accretion disc radius
$R_s = 4$ lt-day is based on extensive work on the subject by a number of
authors \citep{j12,j15,m13,m18}, and results in a very good fit to the
data as may be seen in the left hand panel of Fig.~\ref{fig7}.  The
alternative assumption of a constant mass-to-light ratio for the quasar
disc produces a few outliers from low luminosity quasars, but otherwise
gives a similar fit. In Fig.~\ref{fig8} the data are shown re-plotted as
cumulative histograms in the form originally proposed by \cite{s93}.  The
combined model including intrinsic variations combined with microlensing
amplification from a cosmological distribution of $0.3 M_\odot$ compact
bodies making up the dark matter closely reproduces the observations,
without the need for parameter optimization.  There are a number of strong
constraints on the identity of such dark matter compact bodies which have
been discussed in detail by \cite{h20a}, with the conclusion that the only
plausible candidates are primordial black holes.

The idea that microlensing amplification contributes to the observed
distribution of quasar amplitudes provides a possible
solution to the long standing question of why the amplitude of quasar
variation appears to decline with luminosity, contrary to theoretical
expectations \citep{k04}.  Luminous quasars with larger accretion discs
will become progressively less affected by microlensing amplifications,
thus producing the observed trend.  Some evidence in support of this can
be seen by considering the optical variability of the well known
Palomar-Green sample of nearby bright quasars \citep{g99}.
Fig.~\ref{fig9} shows a plot of amplitude over a 7 year period versus
absolute magnitude $M_B$ for the 42 quasars.  The plot shows no
correlation between amplitude luminosity, with a correlation coefficient
of 0.16.  The maximum redshift of these quasars is $z = 0.371$,
corresponding to an optical depth to microlensing $\tau = 0.010$,
or a probabilty of significant microlensing amplification of less than
1\%.  This provides a consistent explanation for the difference between
the Palomar-Green lack of correlation of amplitude with luminosity, and
the well-known anti-correlation seen in other samples.

\section{Conclusions}

The stimulus for this paper has been the detection of a population of
compact bodies in the Milky Way halo, and the halos of more distant
galaxies and galaxy clusters.  These bodies are most plausibly identified
as primordial black holes, and would make up a large component, and
possibly all, of the dark matter.  In this case these bodies should betray
their presence by the microlensing amplification of quasar lightcurves in
the general field.  The starting point of the investigation was early work
setting limits on a population of stellar mass bodies by considering the
absence of very large amplitude fluctuations in quasar lightcurves.
Later work showed that the adoption of a more realistic size for the
quasar accretion disc removed the expectation of large amplitude quasar
fluctuations in contradiction with observations.  However, the question
still remained as to the extent to which quasar variations in brightness
were amplified by compact bodies.

This paper is based on a large sample of over 1000 quasar light curves,
monitored over a period of 26 years and covering a wide range of
luminosities and redshifts.  The amplitude distribution of luminous
quasars, where the size of the accretion disc is too large to
permit microlensing of stellar mass bodies, was used as a template for the
intrinsic variations in quasar luminosity for the whole sample.  It was
found that to provide a match to the data a significant contribution from
microlensing amplification was required, and so the resulting synthetic
lightcurves were superimposed on a simulated microlensing amplification
pattern for a $\Lambda$CDM Universe with the dark matter made up of
$0.3 M_\odot$ compact bodies.  The input parameters for this procedure,
such as the mass of the lenses and the size of the quasar accretion disc,
were derived from observations and not from optimized parameter fits.  The
resulting distribution of amplitudes after including the effects of
microlensing closely matches that of the observed amplitude distribution
for the quasar sample.  The identity of the lenses is still uncertain, but
the only plausible candidates appear to be stellar mass primordial black
holes.  The overall conclusion of the paper is that to understand the
amplitude distribution of a large sample of quasar light curves it is
necessary to include the microlensing effects of a cosmologically
distributed population of stellar mass compact bodies.

\section*{Data Availability}

The data upon which this paper is based are all publicly available and
are referenced in Section~\ref{obs}.

\bsp	
\label{lastpage}
\end{document}